# Shoulder Surfing attack in graphical password authentication

**ARASH HABIBI LASHKARI**
Computer Science and Data Communication (MCS),
University Malaya (UM)
Kuala Lumpur, Malaysia
a_habibi_l@hotmail.com

**SAMANEH FARMAND**
Computer Science and Information Technology (IT),
University Malaya (UM)
Kuala Lumpur, Malaysia
mobina23@gmail.com

**Dr. OMAR BIN ZAKARIA**
Computer Science and Data Communication (MCS),
University of Malaya (UM), Kuala Lumpur,
Malaysia
omarzakaria@um.edu.my

**DR. ROSLI SALEH**
Computer Science and Data Communication (MCS),
University of Malaya (UM), Kuala Lumpur,
Malaysia
rosli_salleh@utm.edu.my

*Abstract-* Information and computer security is supported largely by passwords which are the principle part of the authentication process. The most common computer authentication method is to use alphanumerical username and password which has significant drawbacks. To overcome the vulnerabilities of traditional methods, visual or graphical password schemes have been developed as possible alternative solutions to text-based scheme. A potential drawback of graphical password schemes is that they are more vulnerable to shoulder surfing than conventional alphanumeric text passwords. When users input their passwords in a public place, they may be at risk of attackers stealing their password. An attacker can capture a password by direct observation or by recording the individual's authentication session. This is referred to as shoulder-surfing and is a known risk, of special concern when authenticating in public places. In this paper we will present a survey on graphical password schemes from 2005 till 2009 which are proposed to be resistant against shoulder surfing attacks.

*Keywords: Graphical Password, Shoulder Surfing, Authentication Scheme, Passwords, Graphical Authentication, Password Attacks.*

## I. INTRODUCTION

Current authentication systems suffer from many weaknesses. The vulnerabilities of the textual password have been well known. Users tend to pick short passwords or passwords that are easy to remember, which makes the passwords unprotected for attackers to break. Furthermore, textual password is vulnerable to guessing, dictionary attack, key-loggers, and social engineering, shoulder-surfing, hidden-camera and spyware attacks. To conquer the limitations of text-based password, techniques such as two-factor authentication and graphical password have been put in use. Other than that, applications and input devices such as mouse, stylus and touch-screen that permit make the appearance of the graphical user authentication techniques possible. However, they are mostly vulnerable to shoulder-surfing as well.

Passwords possess many useful properties as well as widespread legacy deployment; consequently we can expect their use for the foreseeable future. Unfortunately, today's standard methods for password input are subject to a variety of attacks based on observation, from casual eavesdropping (shoulder surfing), to more exotic methods. Shoulder-surfing attack occurs when using direct observation techniques, such as looking over someone's shoulder, to get passwords, PINs and other sensitive personal information. As well as when a user enters information using a keyboard, mouse, touch screen or any traditional input device, a malicious observer may be able to acquire the user's password credentials. This is a problem that has been difficult to overcome

## II. SHOULDER SURFING

In this part, we explain sixteen articles from shoulder surfing section of graphical password by focusing on problems, solutions, findings and their future work.
Problem 1: The most common computer authentication method is to use alphanumerical usernames and passwords. This method has been shown to have significant drawbacks. For example, users tend to pick passwords that can be easily guessed. On the other hand, if a password is hard to guess, then it is often hard to remember [1].

**Methodology used:** To address this problem, some researchers have developed authentication methods that use pictures as passwords. The past decade has seen a growing interest in using graphical passwords as an alternative to the traditional text-based passwords. In this paper, they conducted a comprehensive survey of the existing graphical password techniques till 2005. They classified these techniques into two categories: recognition-based and recall based approaches. They discussed the strengths and limitations of each method and pointed out the future research directions in this area. They also tried to answer two important questions: "Are graphical passwords as secure as text-based passwords?"; "What are the major design and implementation issues for graphical passwords?" This survey will be useful for information security researchers and practitioners who are interested in finding an alternative to text-based authentication methods [1].

**Findings/Outcome:** A comparison of current graphical password techniques was presented. Although the main argument for graphical passwords is that people are better at memorizing graphical passwords than text-based passwords, the existing user studies are very limited and there is not yet convincing evidence to support this argument. Their preliminary analysis suggests that it is more difficult to break graphical passwords using the traditional attack methods such as brute force search, dictionary attack or spyware. However, since there is not yet wide deployment of graphical password systems, the vulnerabilities of graphical passwords are still not fully understood. Overall, the current graphical password techniques are still immature. Much more research and user studies are needed for graphical password techniques to achieve higher levels of maturity and usefulness [1].

**Problem 2:** To overcome the shoulder-surfing attack issue without adding any extra complexity into the authentication procedure [2].

**Methodology used:** In line with the recent call for technology on Image Based Authentication (IBA) in JPEG committee, they presented a novel graphical password design in this paper. It rests on the human cognitive ability of association-based memorization to make the authentication more user-friendly, comparing with traditional textual password. Based on the principle of zero-knowledge proof protocol, they further improved their primary scheme to overcome the shoulder-surfing attack issue without adding any extra complexity into the authentication procedure. System performance analysis and comparisons were presented to support their proposals [2].

**Problem 3:** The advantages of pass-thought over many of the existing authentication technologies include changeability, shoulder surfing resistance, and protection against theft and user non-compliance. Disadvantages of pass-thought authentication include the requirement for a new hardware component (including electrodes) to record the user's brain signals, and the associated performance. For this reason, a pass-thought system may not be accepted for widespread use, but perhaps for high-value or high-importance applications or environments (e.g. within banks and governments) [3].

**Methodology used:** Recent advances in Brain-Computer Interface (BCI) technology indicate that there is potential for a new type of human-computer interaction: a user transmitting thoughts directly to a computer. BCI technology to date has been focused on interpreting brain signals for communication and control for the disabled. The BCI requirements of a pass-thought system are entirely different: they require no interpretation of the brain signals, but the use of as much signal information as possible [3].

The presented novel idea for user authentication called pass-thoughts, whereby a user authenticates to a device by "transmitting" a thought. This transmission would occur through a Brain Computer Interface (BCI), tailored specifically for this purpose. The goal of a pass-thought system would be to extract as much entropy as possible from a user's brain signals upon "transmitting" a thought which has the opposite goal from the filtering and many-to-one signal translation that must occur for interpretation of brain signals. Provided that these brain signals can be recorded and processed in an accurate and repeatable way, a pass-thought system might provide a quasi two-factor, changeable, authentication method resistant to shoulder-surfing. The potential size of the space of a pass-thought system would seem to be unbounded in theory, although in practice it will be finite due to system constraints. In this paper, they discussed the motivation and potential of pass-thought authentication, the status quo of BCI technology, and outline the design of what they believed to be a currently feasible pass-thought system. They also briefly mention the need for general exploration and open debate regarding ethical considerations for such technologies [3].

**Findings/Outcome:** There are many unknowns to resolve before pass-thoughts might become the method they envisioned. It is a hope that this idea for a pass-thought system will inspire research into the area of signal processing and translation algorithms that retain as much repeatable information as possible. If the recording and processing of brain signals can be accurate and repeatable, pass-thoughts might become a viable and useful new form of authentication [3].

**Problem 4:** An attacker can capture a password by direct observation or by recording the individual's authentication session while inserting passwords in public. This is referred to as shoulder-surfing [4].

**Methodology used:** Until recently, the only defence against shoulder-surfing has been vigilance on the part of the user. This paper reports on the design and evaluation of a game-like graphical method of authentication that extends the challenge response paradigm to resist shoulder-surfing. In doing so it aims to motivate the user with a fun, game-like visual environment designed to develop positive user affect and counterbalance the drawback of the longer time to input the password [4].

The Convex Hull Click (CHC) Scheme is an effort to develop security innovations with people in mind. As such, it is an example of "usable security," an approach to design of security systems that is gaining increasing attention. This scheme allows a user to prove knowledge of the graphical password safely in an insecure location because users never have to click directly on their password images [4].

**Findings/Outcome:** Usability testing of the CHC scheme showed that novice users were able to enter their graphical password accurately and to remember it over time. However, the protection against shoulder-surfing comes at the price of longer time to carry out the authentication. The user study and interviews supported the overall concept but identified areas of improvement needed to enhance usability and reduce risks [4].

Contextual changes have to do with how the user thinks about the system. Most of the novice users felt the time was acceptable, although it was objectively long compared to a traditional alphanumeric password. Factors that potentially increase its acceptability to users are multiple: high security which warrants taking more time to login, use of CHC in contexts that do not entail logging in at frequent intervals, ease of remembering the pass-icons and inputting the password accurately, and importantly the "fun factor" of a game-like environment [4].

**Future Work:** Future work should target increasing the speed of input of the password. There is no single solution to this problem. Instead, several incremental changes, human, technical, visual, and contextual, will improve the system. Humans can speed up with practice, the system can be tweaked to improve efficiency, and the icons can be improved [4].

Further directions for CHC are to improve the current icons, create additional icon sets, make the security settings more fully realistic, and then test it in a longitudinal study of everyday use. This longitudinal study could be carried out in a research or teaching lab where users log in to computers daily. Investigating the entropy issue of pass-icons and study in more depth the motivational aspects of the game-like approach were also planned [4].

**Problem 5:** Previous research has found graphical passwords to be more memorable than non-dictionary or "strong" alphanumeric passwords. Participants in a prior study expressed concerns that this increase in memorability could also lead to an increased susceptibility of graphical passwords to shoulder-surfing. The seminal question still remains: Can we have both usable and secure authentication systems? In particular, are graphical passwords the leading candidates to address this long-standing challenge, or do the very characteristics that make graphical passwords more memorable and usable lead to increased security vulnerabilities like shoulder-surfing? [5]

**Methodology used:** This appears to be yet another example of the classic trade-off between usability and security for authentication systems. This paper explores whether graphical passwords' increased memorability necessarily leads to risks of shoulder-surfing. To date, there are no studies examining the vulnerability of graphical versus alphanumeric passwords to shoulder-surfing [5].

This paper examined the real and perceived vulnerability to shoulder-surfing of two configurations of a graphical password, Passfaces, compared to non-dictionary and dictionary passwords. A laboratory experiment with 20 participants asked them to try to shoulder surf the two configurations of Passfaces (mouse versus keyboard data entry) and strong and weak passwords. Data gathered included the vulnerability of the four authentication system configurations to shoulder-surfing and study participants' perceptions concerning the same vulnerability. An analysis of these data compared the relative vulnerability of each of the four configurations to shoulder-surfing and also compared study participants' real and perceived success in shoulder-surfing each of the configurations. Further analysis examined the relationship between study participants' real and perceived success in shoulder-surfing and determined whether there were significant differences in the vulnerability of the four authentication configurations to shoulder-surfing [5].

**Findings/Outcome:** Findings indicate that configuring data entry for Passfaces through a keyboard is the most effective deterrent to shoulder-surfing in a laboratory setting and the participants' perceptions were consistent with that result. While study participants believed that Passfaces with mouse data entry would be most vulnerable to shoulder-surfing attacks, the empirical results found that strong passwords were actually more vulnerable [5].

Despite the common belief that non-dictionary passwords are the most secure type of password-based authentication; the results demonstrate that it is in fact the most vulnerable configuration to shoulder-surfing. This result is unexpected, but possibly explainable. A major finding from the study is that secure and usable authentication might be possible when considering shoulder-surfing risks, but that configuration for data entry (i.e., mouse versus numeric keypad) is an important consideration for graphical passwords like Passfaces. Finally, these findings call into question the notion that non-dictionary passwords are universally "better" than dictionary passwords. The risk mitigation from password choice clearly depends on the nature of the attack [5].

**Future Work:** The non-dictionary passwords, being highly vulnerable to shoulder-surfing attacks is a finding that calls for further investigation. Future studies may investigate shoulder-surfing methods used by real hackers (for example multiple cameras or other equipment) as well as investigation of circumstances for most popular shoulder-surfing environments (work, public access points, etc.) Moreover, further studies may focus on typing speed and possible training effects from long-term use of passwords (both dictionary and non-dictionary) to better establish the impact of long-term use of passwords on their shoulder-surfing vulnerability [5].

**Problem 6:** A potential drawback of graphical password schemes is that they are more vulnerable to shoulder surfing than conventional alphanumeric text passwords [6].

**Methodology used:** They presented a variation of the Draw-a-Secret scheme originally proposed by Jermyn et al. that is more resistant to shoulder surfing through the use of a qualitative mapping between user strokes and the password, and the use of dynamic grids to both obfuscate attributes of the user secret and encourage them to use different surface realizations of the secret. The use of qualitative spatial relations relaxes the tight constraints on the reconstruction of a secret; allowing a range of deviations from the original. They described QDAS (*Qualitative Draw-A-Secret*), an initial implementation of this graphical password scheme, and the results of an empirical study in which they examined the memorability of secrets, and their susceptibility to shoulder-surfing attacks, for both *Draw-A-Secret* and QDAS [6].

**Findings/Outcome:** In their preliminary empirical study QDAS proved to be more resistant to shoulder surfing than its DAS counterpart [6].

**Future Work:** In future they planned to further analyze QDAS by running more studies, and in particular they hope to accurately simulate the context of shoulder-surfing scenario to improve the ecological validity of their findings [6].

**Problem 7:** Shoulder-surfing is a problem that has been difficult to overcome [7].

**Methodology used:** An EyePassword, a system that mitigates the issues of shoulder surfing via a novel approach to user input was presented which is an alternative approach to password entry, based on gaze that deters or prevents a wide range of these attacks. They demonstrated through user studies that their approach requires marginal additional entry time and has accuracy similar to traditional keyboard input, while providing an experience preferred by a majority of users. With EyePassword, a user enters sensitive input (password, PIN, etc.) by selecting from an on-screen keyboard using only the orientation of their pupils (i.e. the position of their gaze on screen), making eavesdropping by a malicious observer largely impractical. They presented a number of design choices and discussed their effect on usability and security. They conducted user studies to evaluate the speed, accuracy and user acceptance of their approach [7].

**Findings/Outcome:** Results demonstrated that gaze-based password entry requires marginal additional time over using a keyboard, error rates are similar to those of using a keyboard and subjects preferred the gaze-based password entry approach over traditional methods [7].

A password can be strengthening by extracting a few additional entropy bits from the gaze path that the user follows while entering the password. Supposedly, the user will follow a similar path, with similar dwell times, every time. A different user, however, may use completely different dwell times. As a result, stealing the user's password is insufficient for logging in and the attacker must also mimic the user's gaze path. A similar technique was previously used successfully to enhance the entropy of passwords entered on a keyboard. While their results showed that the trigger-based mechanism had considerably higher error rates due to eye-hand coordination, it is conceivable that this can be accounted for algorithmically by examining the historical gaze pattern and correlating it with trigger presses [7].

**Problem 8:** To gain access to computer systems, users are required to be authenticated. This is usually accomplished by having the user enter an alphanumeric username and password. Users are usually required to remember multiple passwords for different systems and this poses such problems as usability, memorability and security. Passwords are usually difficult to remember and users have developed their own methods some of which are not secure of selecting passwords which are easy to remember. The main weakness of graphical password systems is shoulder surfing [8].

**Methodology used:** In this research a secure and usable password system which addresses the memorability problem was developed. ToonPasswords is an alternative to traditional text passwords. It draws on the best usability features of existing systems, but provides enhanced security. It reduces the memory load on students by giving them familiar cartoon characters which are demonstrate and are easier to recall than a typical secure text password. Unlike some systems these images are system generated. This avoids users selecting images which might be familiar to an attacker who knows the user personally. They increased the number of images on a screen thus making the probability of a lucky guess as low as 1/64,000. They locked the user out after ten attempts to thwart the most determined and patient of attackers. Giving the user up to ten chances should alleviate frustration when an incorrect password is guessed since the user has more chances. With ToonPasswords the problem of shoulder surfing was overcame by randomizing the location of images at each login [8].

**Findings/Outcome:** This system was shown to be secure based on the probability of guessing a password and on the likelihood of an observer "shoulder surfing" the password and on the difficulty of launching a brute force attack against a graphical image system. Their work demonstrated that security and usability can be achieved simultaneously. It lays the foundation for developing a class of similar password systems, differing only in the degree of security required. Their password system with its low memory requirements can be used in a wide array of applications [8].

**Future Work:** For future work the proposed password system will be implemented and tested for security and usability with real users. Eventually the size of the grid will be increased and more screens will be added to offer more security. ToonPasswords will be compared with text passwords and eventually they want to implement the system on mobile devices [8].

**Problem 9:** Textual password is vulnerable to shoulder-surfing, hidden-camera and spyware attacks. Graphical password schemes have been proposed as a possible alternative to text-based scheme. However, they are mostly vulnerable to shoulder-surfing as well [9].

**Methodology used:** In this paper, they proposed a Scalable Shoulder-Surfing Resistant Textual-Graphical Password Authentication Scheme (S3PAS). This model seamlessly integrates both graphical and textual password schemes and provides nearly perfect resistant to shoulder-surfing, hidden-camera and spyware attacks. It can replace or coexist with conventional textual password systems without changing existing user password profiles. Moreover, it is immune to brute-force attacks through dynamic and volatile session passwords. S3PAS can accommodate various lengths of textual passwords, which requires zero-efforts for users to migrate their existing passwords to S3PAS. Further enhancements of S3PAS scheme are proposed and briefly discussed. Theoretical analysis of the security level using S3PAS is also investigated [9].

**Findings/Outcome:** However, there are still some minor drawbacks in this system similar to other graphical password schemes. The major issues in S3PAS schemes include slightly more complicated and longer login processes. They planned to design a simplified version of S3PAS with a little lower security level to ease its adoption [9].

**Problem 10:** Previous efforts involving picture-based passwords have not focused on maintaining a measurably high level of entropy. Since password systems usually allow user selection of passwords, their true entropy remains unknown [10].

**Methodology used:** A protocol for ignoring duplicate inputs was presented. A shoulder-surfing resistant input method was also evaluated, with six out of 15 users performing an insecure behaviour. A 23-participant study was performed in which picture and character-based passwords of equal strength were randomly assigned. Memorability was tested with up to one week between sessions [10].

In this case, the picture-password group performed better than the character group after one week: 100% recall and 67% respectively, with all 15 picture-group participants correctly selecting only their password items within two tries. This appeared to be a confirmation of the picture superiority effect, and may also be attributable to the "multiple encodings" of each password item (each item was represented by a picture as well as a keyboard key and location in the home grid). However, when ordered passwords with a full 50 bits of entropy were considered, performance for both picture and character passwords was quite poor: 67% recall and 50% respectively. Serial order information either does not benefit from the multiple encodings of picture-password items or passwords at this entropy level are too difficult to remember. A couple observations about user behaviour were also made. Most importantly, the fact that users repeat incorrect inputs is likely based on the fact that users do not receive adequate feedback when entering a password (they cannot see the actual text submitted to the system). Since this is unavoidable for security reasons, the duplicate inputs should be discarded by the authentication server and not counted against the user. This does not compromise the security of the system, since attackers have nothing to gain from duplicating inputs. User behaviour during the SSR task was unexpected. The purpose of the task had been explained immediately before it was performed, yet six out of 15 participants revealed their password through an insecure behaviour. This highlights the importance of usability testing in security applications [10].

**Findings/Outcome:** The study found that both character and picture passwords of very high entropy were easily forgotten. However, when password inputs were analyzed to determine the source of input errors, serial ordering was found to be the main cause of failure. This supports a hypothesis stating that picture-password systems which do not require ordered input may produce memorable, high-entropy passwords. Input analysis produced another interesting result, that incorrect inputs by users are often duplicated. This reduces the number of distinct guesses users can make when authentication systems lock out users after a number of failed logins. Across all conditions, picture passwords were more memorable than character passwords, though the difference was not significant due to the small sample size of the study. It was marginally significant when input data was analyzed to determine how well participants would have performed at an unordered input task [10].

**Future Work:** Picture passwords are a relatively new area of study, so the possibilities for future work are extensive. Based on the results presented here, the most promising future work is in the area of unordered, randomly-assigned passwords. Research into insecure user behaviours and training methods is also extremely important [10].

**Problem 11:** One common practice in relation to alphanumeric passwords is to write them down or share them with a trusted friend or colleague. Graphical password schemes often claim the advantage that they are significantly more secure with respect to both verbal disclosure and writing down. In this paper they investigated the reality of this claim in relation to the Passfaces graphical password scheme [11].

**Methodology used:** By collecting a corpus of naturalistic descriptions of a set of 45 faces, they explored participants' ability to associate descriptions with faces across three conditions in which the decoy faces were selected: (1) at random; (2) on the basis of their visual similarity to the target face; and (3) on the basis of the similarity of the verbal descriptions of the decoy faces to the target face [11].

They conducted an informal pilot study using their implementation of the passpoints system. 5 users ability to select click points in response to verbal descriptions were investigated. The experiment involved the listener sitting in front of a computer screen with the passpoints software loaded. A male experiment moderator stood behind the participant and described the location of each click point in turn. No gesticulation was allowed and descriptions were not permitted to include reference to the current position of the mouse pointer, only points on the image. The listener was allowed to request more information or clarification to which the describer could respond. The results of this pilot study showed 4 out of 5 participants were able to correctly interpret the descriptions into the correct sequence of click points. Participants were found to perform significantly worse when presented with visual and verbally grouped decoys, suggesting that Passfaces can be further secured for description. Subtle differences in both the nature of male and female descriptions, and male and female performance were also observed [11].

**Findings/Outcome:** This study has in part demonstrated the degree to which Passfaces can be verbally described, but also how through judicious choice of decoys we can reduce the vulnerability of Passfaces to description. Moreover, this empirical study has highlighted the reality that contrary to common wisdom users can share Passfaces graphical passwords. They also anticipated that the vulnerability of graphical password schemes to description could have impact on issues of both memorability and shoulder surfing. If decoy faces in Passfaces employed a grouping policy that means that they were in some way similar to the target face, it is likely to impact on the ability of a user to select the target face. Likewise, in a shoulder-surfing scenario, attackers have a limited amount of time to form a quick memory association with components of an authentication secret [11].

**Problem 12:** User authentication is one of the important topics in information security. Traditional strong password schemes could provide with certain degree of security; however, the fact that strong passwords being difficult to memorize often leads their owners to write them down on papers or even save them in a computer file. As a result, security becomes greatly compromised [12].

**Methodology used:** On the other hand, knowing that human beings are predominant visual creatures, many researchers have investigated or developed graphical password schemes recently. In this paper, they proposed a graphical password scheme for user authentication using images with random tracks of geometric shapes. This method is not only more secure than most of the existing graphical password schemes, it also solves problems like requiring a large image database, uneasy to repeat mouse clicking at the same position, as well as images being too simple to cause collisions on points selected for different users [12].

**Findings/Outcome:** They justified that graphical password random geometric graphical password (RGGPW) indeed is robust against common security attacks like brute-force search, spyware, shoulder surfing, social engineering, and forgery. They also showed the images to demonstrate user friendliness in both recognition and selection of pass-objects from the given images. In addition, RGGPW is storage-efficient as all images are created when needed [12].

**Future Work:** Working on how to create images with more complex tracks and easier recognizable objects and implementing a website to test the acceptance of this technique is in their future work [12].

**Problem 13:** Alphanumeric passwords are widely used in computer and network authentication to protect users' privacy. However, it is well known that long, text-based passwords are hard for people to remember, while shorter ones are susceptible to attack [13].

**Methodology used:** Graphical password is a promising solution to this problem. Draw-A-Secret (DAS) is a typical implementation based on the user drawing on a grid canvas. Currently, too many constraints result in reduction in user experience and prevent its popularity. A novel graphical password strategy Yet Another Graphical Password (YAGP) inspired by DAS is proposed in this paper. In a 48×64 grid, the secret drawings can be described in detail. The users can concentrate on the drawing to improve user experience because exact positions are not required in YAGP. Meanwhile, the algorithm proposed in YAGP is trend-sensitive which actually reflects drawing trends. The proposal has the advantages of free drawing positions, strong shoulder surfing resistance and large password space. Experiments illustrated the effectiveness of YAGP. Furthermore, user personalities have a great influence on the drawings and therefore make it harder for others to imitate. Additionally, users can draw the secrets small enough to resist shoulder surfing [13].

**Findings/Outcome:** Some preliminary experiments are carried out. The results showed that YAGP achieves an encouraging performance in usability and security and possesses a high resistance to shoulder surfing [13].

The main drawback of YAGP is that it's hard to redraw the password precisely. The legal user cannot always be assured to login successfully because the gaps between user drawings are uncertain while the similarity threshold value is fixed [13].

**Future Work:** Future research will concentrate on improving YAGP as well as developing a comparison algorithm of higher efficiency in distinguishing the legal user from attackers [13].

**Problem 14:** Threats such as key-loggers, weak password, and shoulder surfing, forcing the user to memorize different passwords or carrying around different tokens, "familiarization" or a lengthy "password setup" process are today's drawbacks of authentication systems [14].

**Methodology used:** In this paper, they proposed a new authentication scheme based on graphical password and multifactor authentication. To that end, they employed the user's personal handheld device as the password decoder and the second factor of authentication. In these methods, a service provider challenges the user with an image password. To determine the appropriate click points and their order, the user needs some hint information transmitted only to her handheld device. They showed that this method can overcome threats such as key-loggers, weak password, and shoulder surfing [14].

Their approach can be effectively and securely used as user-friendly authentication mechanism for public and un-trusted terminals. The proposed solution is unique in many ways: 1. It is the first graphical password solution that employs two-factor authentication. 2. They never assume the handheld device is trusted. 3. This solution resists screen recording attacks. 4. This method doesn't need a "familiarization" or a lengthy "password setup" process. 5. Lost or stolen handheld doesn't expose a security risk [14].

**Findings/Outcome:** With the increasing popularity of handheld devices such as cell phones, this approach can be leveraged by many organizations without forcing the user to memorize different passwords or carrying around different tokens. This system can be applied to more than just authentication mechanisms: their system is applicable anywhere that there is a need to enter sensitive or private data. For instance, Social Security Number can be entered via this system without leaking or revealing any directly usable information to the terminal or even the handheld device [14].

**Problem 15:** Two Factor authentication mechanisms are considered to be secure for authenticating a user in Internet based environment. As the number of services provided online is day by day increasing, users intending to use various online services are also increasing. With each service requiring the user to register separately, the overhead of remembering many ID/password pairs has lead to the problem of memorability. To address this, researchers have proposed mechanisms for multi-server environment where in the user needs to register with a single registration centre using one ID/password pair and thereby access all the services registered through that server. But, as these mechanisms employ textual passwords, they suffer from many inherent drawbacks [15].

**Methodology used:** In this paper they proposed a two factor password authenticated key agreement mechanism using graphical password where in the user needs to recognise his secret image presented to him as challenge. The scheme allows the user to choose images from a given set of image categories. The scheme is based on simple collision resistant hash functions. The protocol was designed such that it does not maintain verification table at the server for authentication and to resist replay attack it does not employ the time concurrency mechanism which has various weaknesses. Instead, it uses random nonce. In addition, the protocol provides secure low computation mutual authentication and session key agreement [15].

**Findings\Outcome:** It is secure against several ID thefts, Insider attack, Replay attack, Shoulder surfing attack, Reconnaissance attack, Server spoofing attack and guessing attack. The proposed scheme will work efficiently on wired network scenario. Their system is secure, efficient and user friendly authentication scheme which has several unique features [15].

**Future Work:** They intend to explore further, the same concept and protocol for the wireless domain where there are bandwidth constraints [15].

**Problem 16:** To overcome the vulnerabilities of traditional methods, visual or graphical password schemes have been developed. Because simply adopting graphical password authentication also has some drawbacks, some hybrid schemes based on graphic and text were developed [16].

**Methodology Used:** In this paper, a stroke-based textual password authentication scheme was proposed. It uses shapes of strokes on the grid as the origin passwords and allows users to login with text passwords via traditional input devices. The method provides strong resistant to hidden-camera and shoulder-surfing. Moreover, the scheme has flexible enhancements to secure the authentication process. The analysis of the security of this approach was also discussed [16].

**Findings/Outcome:** The scheme has salient features as a secure system for authentication immune to shoulder-surfing, hidden camera and brute force attacks. It also has variants to strengthen to security level through changing the login interface of the system [16].

However, the system still has some drawbacks. Firstly, this method is relativity unfamiliar to the general people so that the users may adopt the simple and weak strokes as their passwords. Secondly, the process of creating original password is more vulnerable than the login step. Thirdly, the login process is longer than other graphical schemes [16].

**Future Work:** Designing more advanced authentication system to improve this method [16].

### III. COMPARISON TABLE

In our study, we found twelve methodologies which overcome shoulder surfing attack in graphical password. We tried to highlight them in a table and demonstrated drawbacks and future works of each scheme briefly. The below Table indicates the result of our study on sixteen articles on shoulder surfing.

TABLE 1: COMPARISON TABLE ON SHOULDER SURFING ATTACK METHODOLOGIES

| Problem | Methodology | Drawbacks/Future Work |
|---|---|---|
| Shoulder- Surfing Attack | association-based memorization, zero-knowledge proof protocol [2] | N/A |
| | signal processing and translation algorithms, extract as much entropy as possible from a user's brain signals [3] | not accurate and repeatable recording and processing of brain signals, requirement for a new hardware component and the associated performance |
| | game-like graphical method of authentication that extends the challenge response paradigm [4] | longer time to carry out the authentication |
| | qualitative mapping between user strokes and the password, use of dynamic grids to both obfuscate attributes of the user secret, use different surface realizations of the secret [6] | N/A |
| | user enters sensitive input by selecting from an on-screen keyboard using only the orientation of their pupils[7] | similar error rates to those of using a keyboard and needs marginal additional time over using it |
| | randomizing the location of images(familiar system generated cartoon characters)at each login [8] | increasing the size of the grid, adding more screens to offer more security |
| | integrates both graphical and textual password schemes without changing existing user password profiles [9] | Slightly more complicated and longer login processes. |
| | using images with random tracks of geometric shapes for authentication [12] | creating images with more complex tracks and easier recognizable objects, implementing a website to test acceptance of this technique |
| | inspired by DAS, algorithm reflects drawing trends, free drawing positions, users can draw the secrets small enough, users personalities have a great influence on the drawings [13] | hard to redraw the password precisely, the gaps between user drawings are uncertain while the similarity threshold value is fixed |
| | using handheld device as a password decoder and sending hint information to determine appropriate click points and their order, second factor of authentication [14] | N/A |
| | images are given by the system, simple collision resistant hash functions, no verification table at the server for authentication [15] | N/A |
| | using shapes of strokes on the grid as the origin passwords, changing the login interface of the system [16] | method relativity unfamiliar to the general people, longer login process and more secure than registration phase |

## IV. CONCLUSION

In this paper we studied on more than 30 graphical password designs then selected 16 algorithms resistant to shoulder surfing. We emphasized the problem, methodology and future works of them and brought out a summery table of our work at the end.

To have a good system high security and good usability are both needed and cannot be separated. Shoulder surfing attack is under security provision. There are few proposed methods to shoulder surfing problem but they still need to be improved.

In our next paper in graphical password algorithms, we will propose an enhancement on one of graphical password algorithm in recognition-based category to solve limitations of graphical password scheme.

We hope this study be useful for those who have new ideas on secure and useable graphical authentication system.

ACKNOWLEDGMENT